\documentstyle[aps,epsf,multicol,prl]{revtex}
\begin{document}

\draft
\title{Low Temperature Phase Transition in
Sr$_{0.66}$Ba$_{0.34}$Nb$_2$O$_6$ Single Crystal Fibers}
\author{J. L. B. Faria,\thanks{Author to whom
correspondence should be adressed: e-mail: hulk@fisica.ufc.br} C.
W. A. Paschoal, P. T. C. Freire, A. P. Ayala,
\\ F. E. A. Melo, J. Mendes Filho,}
\address{Departamento de F\'{\i}sica, Universidade Federal do
Cear\'{a}, Caixa Postal 6030,\\  60455-760 Fortaleza, Cear\'{a},
Brazil}
\author{I. A. Santos, J. A. Eiras}
\address{Departamento de F\'{\i}sica, Universidade Federal de S\~ao Carlos,
Caixa Postal 676,\\ 13565-670 S\~ao Carlos, S\~ao Paulo, Brazil}

\date{\today}
\pagebreak
\maketitle

\begin{abstract}
The structural changes in Sr$_{0.66}$Ba$_{0.34}$Nb$_2$O$_6$
single crystal fibers when the temperature decreases from 295 to
10 K is investigated by dielectric constant measurements and
Raman spectroscopy. The anomaly observed in the plot of
$\varepsilon"$ is associated to the variation in the intensity
and frequency of some Raman modes. We observe that the intensity
of some low energy modes has a very singular behavior for the
$y(xz)\overline{y}$ scattering geometry. In the high energy
region ($>$ 850 cm$^{-1}$), a band at $\sim$ 880 cm$^{-1}$
disappears when the temperature is cooled down to 10K for the
$y(xz)\overline{y}$ geometry and shifts to higher energy values
for the $y(zz)\overline{y}$ geometry.
\end{abstract}

\begin{multicols}{2}[]

\section{Introduction}

Strontium barium niobate Sr$_x$Ba$_{1-x}$Nb$_2$O$_6$ (SBN) is a
ferroelectric system with remarkable electro-optic properties.
Thin films have been used in optical data-storage\cite{memoria},
microelectronic devices applications and crystalline and
amorphous substrates\cite{filmes,filmes2}. Beside its
technological applications, SBN exhibits interesting structural
characteristics as incommensurate superstructures\cite{2-6},
polarization memory\cite{1-6} and high stability to intense laser
radiation source\cite{xia-5}. Raman spectroscopy has been employed
to study the variation of the vibrational modes of SBN in
connection to its ferroelectric
characteristics\cite{clarke-11,burns-4}.

From the studies performed in the late 60's, it is known that the
room temperature crystalline structure of SBN belongs to the
"tungsten-bronze" family, whose the main compound is K$_x$WO$_3$.
Since the sites of this structure are not completely filled, a
large variety of compositional formulas is found in the
literature\cite{jamie1}. Compounds of SBN with x ranging from
0.25 up to 0.75 (the range of stability for the solid
solution\cite{solida,solida2}) were studied by Jamieson {\it et.
al.}\cite{jamie1}.

There are some evidences of the existence of one or two phase
transitions in these bulk compounds when the temperature is
lowered down to 10 K\cite{medidascd,22}. The techniques employed
were X-ray diffration, specific heat and pyroelectric coefficient
measurements in these references. Although Raman spectroscopy
study has been used to investigate the ferroelectric-paraelectric
transition in SBN occurring for T $>$ 300 K\cite{wilde-6}, no one
has employed this technique to investigate the phase transitions
occurring for T $<$ 300 K. We address here the phase transition
undergone by SBN (x=0.66) at about 80 K as observed both by Raman
spectroscopy and dielectric constant measurements giving an
additional evidence via optical measurements for a phase
transition undergone by SBN crystal fibers at low temperatures.

\section{experimental}

The SBN samples used in the experiments were prepared using the
standard oxide mixture route. Ceramic powders with nominal
composition (Sr$_{0.61}$Ba$_{0.39}$)Nb$_2$O$_6 $ - SBN 61/39 were
prepared beginning from the mixture of Nb$_2$O$_5 $,
Ba(NO$_3$)$_2$ and SrCO$_3$ in the wanted proportions. The
precursory powders were mixed in a ball milling during 3 hours for
homogenization, being after calcined for solid state reaction
which takes place at 1300 $^o$C during 3 hours. The samples were
compacted to form a disk shape sample. Finally, the disks (with
the dimensions 25 x 12 x 1 mm$^3 $) were fired at a sintering
temperature of 1350~$^o$C during 3.5 hours. Sticks (with 20 x 0.7
x 0.7 mm$^3 $) were cut of the sintering plates for the growth of
the monocrystalines fibers by using the fusion technique with
laser (LHPG - Laser Heated Pedestal Growth). In this technique the
ceramic sticks are used as seeds for the growth of the fibers. The
composition of the fibers of pure SBN nominally indicated that
everybody was faulty in Ba. The analysis were made through an EDX
microprobe in an electronic microscope Zeiss DSM-960, that
revealed fibers with nominal composition Sr/Ba 61/39 present the
value 66/34.

The Raman spectra were recorded with a Jobin Yvon T64000
spectrometer, equipped with a $N_2-$cooled Charge Coupled Device
(CCD) detection system. The slits were set for a \,2 cm$^{-1}$
spectral resolution. The line 514.5 nm of an Argon ion laser was
used as excitation. A Olympus microscope lens with a focal
distance f = 20.5 mm and numeric aperture NA = 0.35 was used to
focus the laser to sample surface. The incident power density was
of the order of 100~W/cm$^2$. Several measurements were performed
in the temperature range investigated and carried out in the
$y(zz)\overline{y}$ and $y(xz)\overline{y}$ scattering
geometries\cite{porto}, where x,y and z are associated to the
crystallographic axis.

\section{Results and Discussion}

\begin{figure}[tbp]
\begin{center}
\leavevmode
 \[
\epsfxsize 3.2in \epsffile{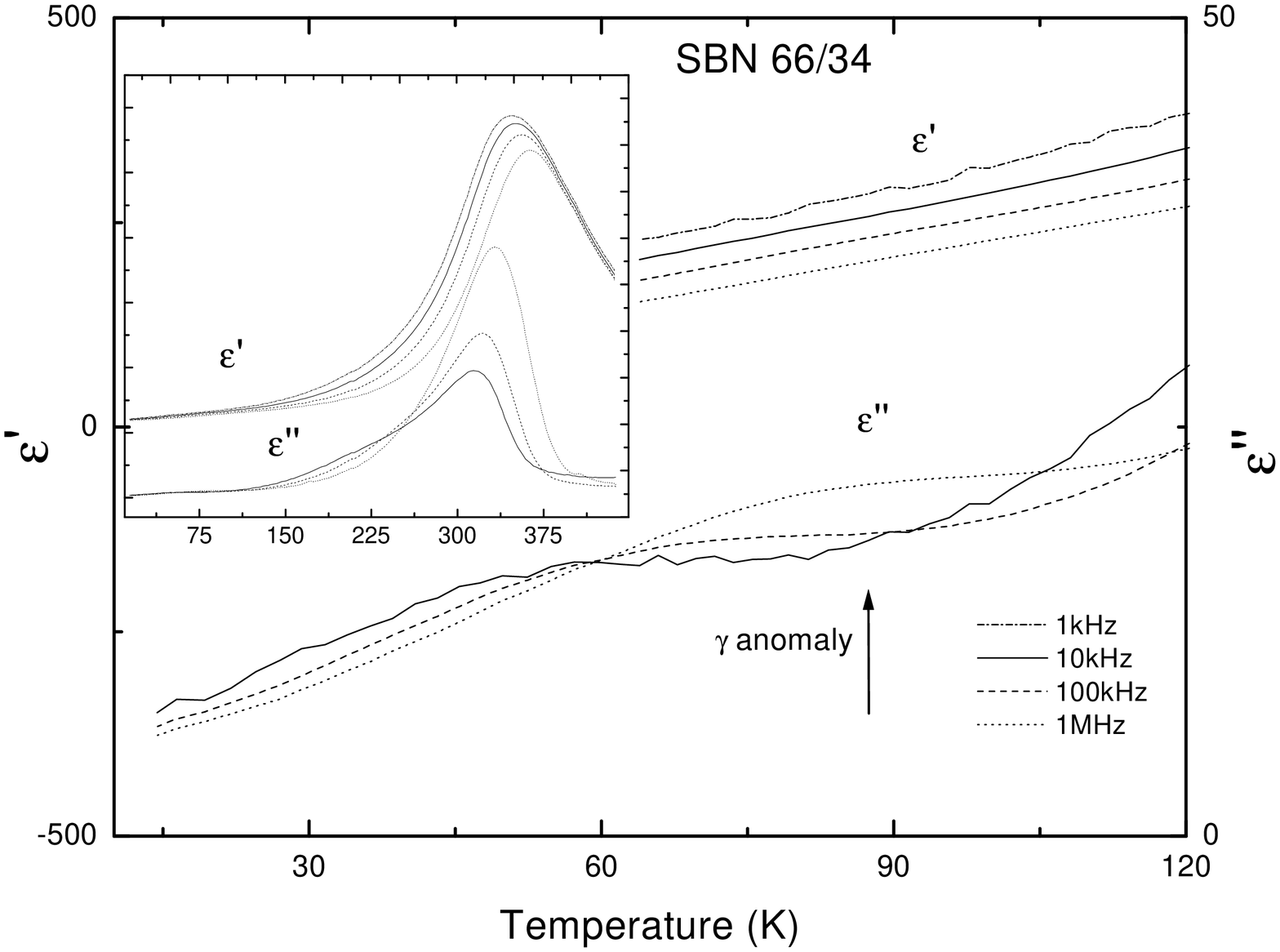}
\]
\end{center}
\caption{Plot temperature {\it vs.} dielectric constants
$\varepsilon'$ and $\varepsilon"$ of the SBN 66/34 for
temperatures between 10 and 450K.} \label{fig-dieletrico}
\end{figure}

The dielectric constant is an ideal parameter to describe the
electric properties of dielectric materials\cite{cd}.  For most
of the materials the dielectric constant in the linear regime is
not a function of the applied electric field intensity, but for
alternate fields, it will depend on frequency. Fig.
\ref{fig-dieletrico} displays both the real ($\varepsilon'$) and
the imaginary ($\varepsilon''$) parts of dielectric constant as a
function of temperature. The dielectric constant of SBN is
characterized by two different features. The most intense is
called $\alpha$ which appears in the inset of
Fig.\ref{fig-dieletrico} for T $\approx$ T$_c$, the Curie
temperature. Such an anomaly has its origin on the temperature and
frequency dependence of the dielectric permittivity of
ferroelectrics with relaxor character near the ferroelectric -
paraelectric phase transition \cite{22}. We will not attempt to
discuss this $\alpha$ - feature because it is well studied
elsewhere \cite{22}. Rather, we will put forward considerations
about another very important characteristic in Fig.
\ref{fig-dieletrico}. The second feature of interest is the
$\gamma$ anomaly which can be associated with polarization
fluctuation originating from the relaxor nature of SBN solid
solution \cite{bursill}. This means that SBN, like
Pb$_{1-x}$Ba$_x$Nb$_2$O$_6$, is a different type of relaxor
ferroelectric where the conventional dipolar model is no valid
anymore \cite{21}. For SBN we must consider both local and
macroscopic polarization states, where the former differ slightly
from the latter. The two possible local states are separated by
an energy barrier in the energy well. At high temperatures these
barriers can be overcome due to thermal energy. However, for low
temperatures the activation barrier is comparable to the thermal
energy allowing relaxation processes through changes in the
polarization states.

Another interpretation for the $\gamma$ anomaly is a diffuse
phase transition undergone by SBN crystal fiber at 80 K. This
interpretation is established by the work of Ref. \cite{medidascd}
where spontaneous polarization, pyroelectric coefficient and
dielectric constant measurements performed on
Sr$_{0.5}$Ba$_{0.5}$Nb$_2$O$_6$ crystal bulk point to a phase
transition between 60 and 80 K. Threrefore, it is plausible to
expect that this anomaly observed in SBN crystal fibers should be
due to a phase transition undergone by the material.

\begin{figure}[tbp]
\begin{center}
\leavevmode
 \[
\epsfxsize 2.0in \epsffile{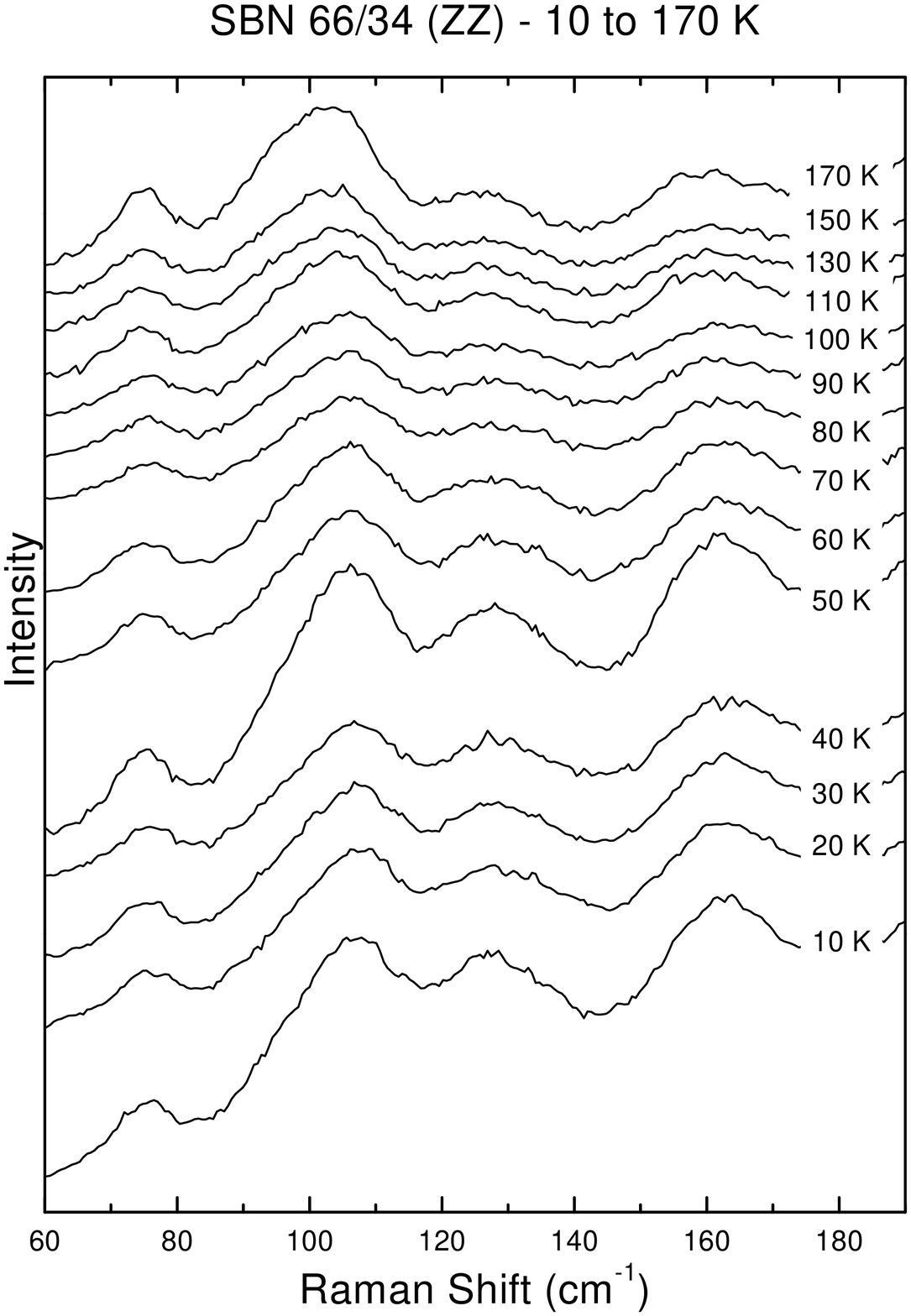}
\]
\end{center}
\caption{Raman spectra of SBN 66/34 in the scattering geometry
$y(zz)\overline{y}$ for temperatures between 10 and 170 K.}
\label{fig-espectrosZZ1reg1}
\end{figure}

\begin{figure}[tbp]
\begin{center}
\leavevmode
 \[
\epsfxsize 2.0in \epsffile{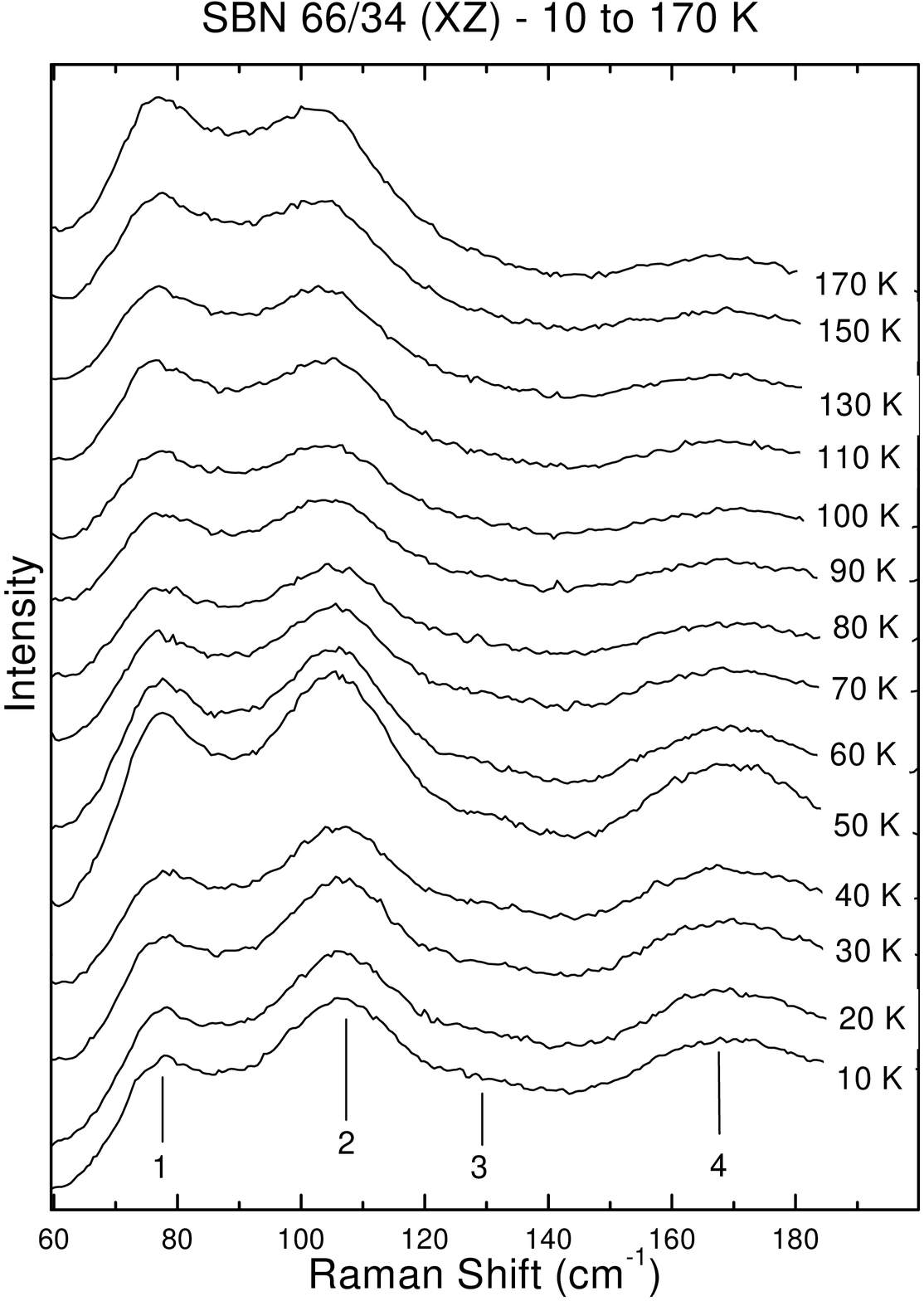}
\]
\end{center}
\caption{Raman spectra of SBN 66/34 in the scattering geometry
$y(xz)\overline{y}$ for temperatures between 10 and 170 K.}
\label{fig-espectrosXZ1reg1}
\end{figure}

\begin{figure}[tbp]
\begin{center}
\leavevmode
\[
\epsfxsize 2.0in \epsffile{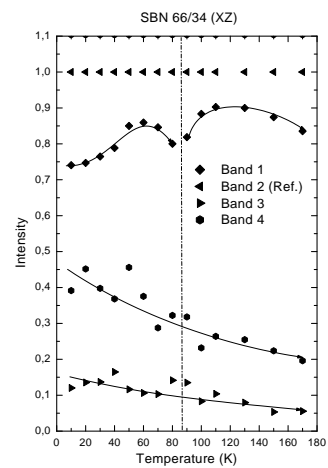}
\]
\end{center}
\caption{Plot intensity normalized {\it vs.} temperature of SBN
66/34 in the scattering geometry $y(xz)\overline{y}$ for
temperatures between 10 and 170 K.} \label{fig-fitsXZamp}
\end{figure}

In Figs. \ref{fig-espectrosZZ1reg1} and \ref{fig-espectrosXZ1reg1}
we display the Raman scattering spectra in the low frequency
region at several temperatures recorded in the
$y(zz)\overline{y}$ and $y(xz)\overline{y}$ scattering
geometries, respectively. At the lowest temperature, 8 bands are
observed. Increasing the temperature, these bands become broader
due to anharmonic effects\cite{anarmonicos}. Apparently, no
change is observed. However, if one normalizes all bands to that
at 106 cm$^{-1}$ (band 2), it can be seen that the intensity of
the band at 75 cm$^{-1}$ (band 1) changes abruptly around 85 K,
as shown in Fig \ref{fig-fitsXZamp}. In this plot the background
was properly subtracted\cite{background,background2}. Here we
arrive to an important point: the sudden intensity change of a
low frequency band. The significance of this variation is
discussed as follows.

The intensity of a phonon results from different factors as, for
example, a deformation potential interaction term \cite{loudon}.
This can be associated to a change of structure when the material
is being subjected to a variation of external condition, e.g.,
temperature or pressure. Changes in the intensities of
low-frequency Raman bands evidencing phase transition by
temperature variation were already reported for a well known
ferroelectric crystal, triglycine sulfate, where a phonon of 60
cm$^{-1}$ was connected with the onset of the phase transition
\cite{silber}. By pressure variations, it was also observed
change of structure through the study of intensity of Raman bands
\cite{bib-bene,bib-teix}. In the first case, abrupt changes of
intensity in several bands in the spectra taken place between
2.03 and 2.24 GPa provided strong evidence that modifications in
the unitary cell of the crystal happened in this pressure
interval. In the second case one striking change in the relative
intensities of bands associated to modes of 41 and 48 cm$^{-1}$
is verified for pressures between 2.2 and 2.3 GPa. An additional
low energy band appears in the Raman spectra evidencing a phase
transition is present\cite{bib-teix}. However, changes in the
relative intensity of Raman bands with variation of any
thermodynamic parameter does not imply, necessarily, a phase
transition has occurred. These variations sometimes may be
associated to the internal reordering of ions in the unit cell
but with no change of structure; one example is given by
LiNa$_3$(SO$_4$)$_2$.6H$_2$O crystal, where great modifications
in the Raman intensities occurring when temperature changes from
12 to 300 K are not associated with any phase transition
\cite{bib-jef}. In the case of SBN crystal fibers here
investigated we believed that the change in the Raman intensities
may be credited to a phase change, because, as was already shown,
measurements of dielectric constant for the SBN crystal bulk
point to the existence of a phase transition \cite{medidascd}.

\begin{figure}[tbp]
\begin{center}
\leavevmode
 \[
\epsfxsize 2.0in \epsffile{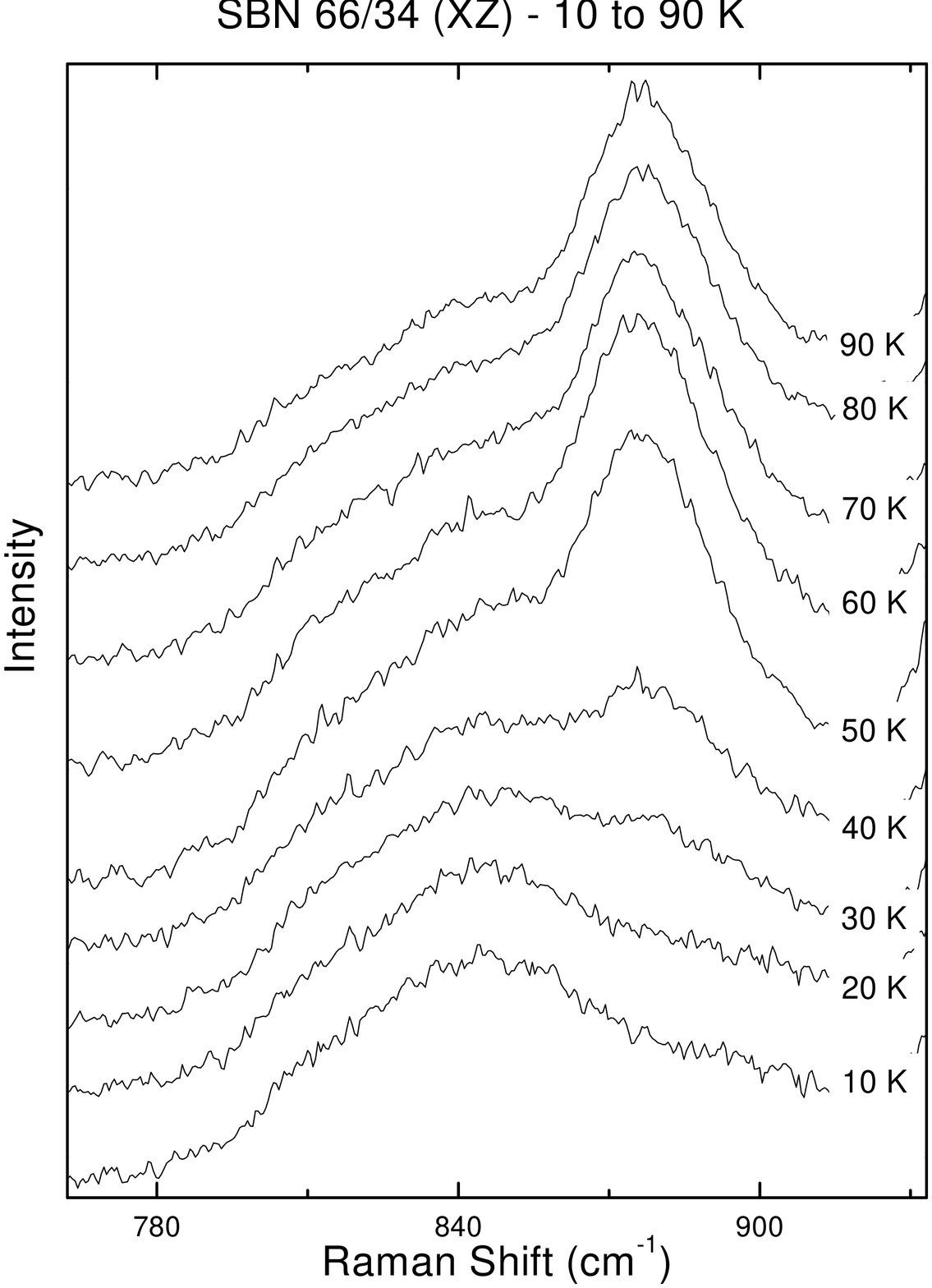}
\]
\end{center}
\caption{Raman spectra of SBN 66/34 in the scattering geometry
$y(xz)\overline{y}$ for temperatures between 10 and 90 K.}
\label{fig-espectrosXZ1reg3a}
\end{figure}

Fig. \ref{fig-espectrosXZ1reg3a} shows the SBN Raman spectra in
the spectral range 760-930 cm$^{-1}$ taken from 10 to 90 K for
the $y(xz)\overline{y}$ scattering geometry. Bands in this region
are due to $\nu_1$ modes of the NbO$_6$ structure
\cite{octaedros}. As observed by other authors \cite{wilde-6}
this region presents an extremely broad band whose rising is
possibly associated to a structural disorder. In the structure
there exist octahedra of NbO$_6$ with the polar direction close
to the O -- Nb -- O axis. These axes are not exactly parallel to
the c axis and, additionally, since we have two non-equivalent Nb
ions, this give rise to two slightly different $\nu_1$ vibrations
of the NbO$_6$ structure. Starting from 90 K, the intensity of
the band at 880 cm$^{-1}$ decreases progressively down to 20 K,
when the band disappears completely. As the origin of these two
bands is associated with structural disorder, this means that for
low temperatures a new configuration is being supplied to [NbO$_6
$]$^{7 -}$ ions. This new configuration may be originated either
by a change in SBN unitary cell as a consequence of the
temperature variation or by a local change of the neighboring
region of the ion.

\begin{figure}[tbp]
\begin{center}
\leavevmode
 \[
\epsfxsize 2.0in \epsffile{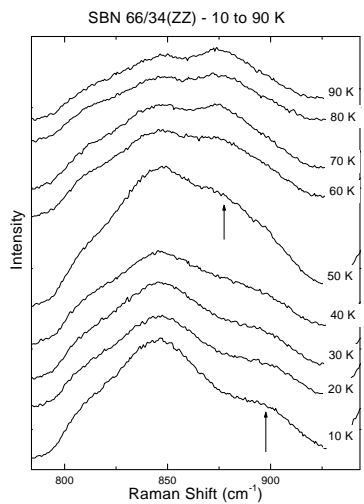}
\]
\end{center}
\caption{Raman spectra of SBN 66/34 in the $y(zz)\overline{y}$
scattering geometry for temperatures between 10 and 90 K. The
arrow indicates the center of the band with the highest energy.}
\label{fig-espectrosZZ1reg3a}
\end{figure}

In Fig. \ref{fig-espectrosZZ1reg3a} we show the $\nu_1$ mode
region for the $y(zz)\overline{y}$ scattering geometry. Two bands
are also observed. Decreasing the temperature, the intensity of
the higher energy band decreases but, instead of disappearing
completely, it remains even at 10 K. When looked carefully around
50 K, the energy of the band with higher Raman shift undergoes a
jump. The arrows placed in spectra of 50 K and 10 K show the
exact value of the center of the bands obtained by lorentzians
fittings. By one hand, the fact that the two bands are present
down to the lowest temperature in this different scattering
geometry means that the two non-equivalent NbO$_6$ structures
remains down to 10 K. On the other hand, a jump in the energy of
one band is pointing to a change of the neighboring around the
octahedral [NbO$_6$]$^{7-}$ ion. Moreover, the fact that some
variation of band intensities of external modes are also being
observed serves as a guide when deciding on the conclusion to be
adopted, in our case, the existence of a structural phase
transition.

As summary, through Raman spectroscopy and dielectric constant
measurements it was possible to verify the existence of a phase
transition for SBN crystal fibers at temperatures below 100 K. The
evidences are: (i) changes in relative intensities of low energy
bands; (ii) disappearance of a high energy band, associated with
the ion [NbO$_6$ ]$^{7 -}$ vibration in the $y(xz)\overline{y}$
scattering geometry and abrupt change in energy of the same phonon
in the $y(zz)\overline{y}$ scattering geometry; (iii) change in
dielectric constant in the same temperature limits where occur the
variations in the low energy modes.

{\bf Acknowledgments}~-~J. L. B. Faria and C.W.A. Paschoal
acknowledge financial support from CNPq and FUNCAP, respectively.
P. T. C. Freire wishes to acknowledge the financial support from
FUNCAP through a grant No . 017/96 P\&D. The authors acknowledge
A.C. Hernandes for the fiber growth and A. G. Souza Filho and I.
Guedes for discussions related to this work. Financial support
from CNPq, FAPESP and FINEP is also gratefully acknowledged.

\end{multicols}
\end{document}